\documentstyle[11pt,paspconf,psfig,epsf]{article}

\begin{document}

\title{MODEL ATMOSPHERES:\\ Brown Dwarfs from the Stellar Perspective}

\author{F. Allard and P. H. Hauschildt}
\affil{Physics Department, Wichita State University, Wichita, KS 67260-0031}
\affil{Department of Physics and Astronomy, University of Georgia, Athens, GA 30602-2451}

\begin{abstract}

In this paper, we review the current theory of very low mass stars
model atmospheres including the coolest known M~dwarfs, M~subdwarfs,
and brown dwarfs, i.e. T$_{\rm eff} \leq 5,000\,$K and $-2.0 \le {\rm
[M/H]} \le +0.0$.  We discuss ongoing efforts to incorporate molecular
and grain opacities in cool stellar spectra, as well as the latest
progress in deriving the effective temperature scale of M~dwarfs.  We
also present the latest results of the models related to the search
for brown dwarfs.

\end{abstract}


\section{Very Low Mass Star models and the T$_{\rm eff}$ Scale}\label{intro}
 
Very Low Mass stars (VLMs) with masses from about $0.3$~M$_{\odot}$ to
the hydrogen burning minimum mass ($0.075$~M$_{\odot}$, Baraffe et al.
1995) and young substellar brown dwarfs share similar atmospheric
properties.  Most of their photospheric hydrogen is locked in H$_2$
and most of the carbon in CO, with the excess oxygen forming important
molecular absorbers such as TiO, VO, FeH and H$_2$O.  They are subject
to an efficient convective mixing often reaching the uppermost layers
of their photosphere.  Their energy distribution is governed by the
millions of absorption lines of TiO and VO in the optical, and H$_2$O
in the infrared, which leave {\bf no} window of true continuum.  But
as brown dwarfs cool with age, they begin to differentiate themselves
with the formation of methane (CH$_4$) in the infrared (Tsuji et
al. 1995) at the expense of CO which bands begin to weaken in their
spectra (Allard et al. 1996).  Across the stellar-to-substellar
boundary, clouds of e.g. corundum (Al$_2$O$_3$), perovskite
(CaTiO$_3$), iron, enstatite (MgSiO$_3$), and forsterite
(Mg$_2$SiO$_4$) may form, depleting the oxygen compounds and heavy
elements and profoundly modifying the thermal structure and opacity of
their photosphere (Sharp \& Huebner 1990, Burrows et al. 1993, Fegley
\& Loggers 1996, Tsuji et al. 1996ab).

Because these processes also occur in the stellar regime where a
greater census of cool dwarfs is currently available for study, a
proper quantitative understanding of VLM stars near the hydrogen
burning limit is a prerequisite to an understanding of the
spectroscopic properties and parameters of brown dwarfs and
jovian-type planets.  Model atmospheres have been constructed by
several investigators over recent years with the primary goals of:

\begin{enumerate}

\item Determining the effective temperature scale of M dwarf stars
      down to the substellar regime.

\item Identifying spectroscopic signatures of substellarity i.e.
      gravity indicators for young brown dwarfs, and spectral features
      distinctive of cooler evolved brown dwarfs.

\item Providing non-grey surface boundary to evolution calculations of
      VLMs and brown dwarfs leading to more consistent stellar models,
      accurate mass-luminosity relations and cooling tracks for these
      objects.

\end{enumerate}

The computation of VLMs and brown dwarf model atmospheres requires a
careful treatment of the convective mixing and the molecular
opacities.  The convection must currently be handled using the mixing
length formalism while a variety of approximations have been used to
handle the millions of molecular and atomic transitions that define
the spectral distributions of VLMs and brown dwarfs.  The most
accurate of these methods is the so-called opacity sampling (OS)
technique which consists in adding the contribution all transitions
absorbing within a selected interval around each point of a
pre-determined wavelength grid (typically $\approx 22000$ points from
0.001 to 100 $\mu$m). When the detail of the list of transitions is
lacking for a molecule as is the case for the important absorber VO,
the Just Overlapping Line Approximation (JOLA) offers an alternative
by approximating the band structure based on only a few molecular
rotational constants.  The straight-mean (SM) and K-coefficients
techniques, which consist in averaging the opacities over fixed
wavelength intervals chosen smaller than the resolution of typical
observations, have also been used in modeling late-type dwarf
atmospheres. Their main advantage is to save computing time during the
calculation of the models, often at the expense of an accurate
spectral resolution.  The list of recent model atmospheres and the
opacity technique they mostly rely upon is given in table~\ref{grids}.

\begin{table}
\caption{Relevant Model Atmospheres}\label{grids}
\begin{center}\scriptsize
\begin{tabular}{lcrc}
\tableline
Authors & Grid & T$_{\rm eff}$ range (K) & Main Opacity Treatment\\
\tableline
\tableline
&&&\\
              Kurucz 1992&          Atlas12 &  $3500 - \dots$~~ &            OS\\
&&&\\
              Allard 1990&             Base &  $2000 - 3750$ &     SM$+$JOLA\\
       Saumon et al. 1994& zero-metallicity &  $1000 - 5000$ &            OS\\
        Tsuji et al. 1995&        grainless &  $1000 - 2800$ &          JOLA\\
               Brett 1995&            MARCS &  $2400 - 4000$ &            OS\\
Allard \& Hauschildt 1995&    Extended Base &  $1500 - 4500$ &            SM\\
        Tsuji et al. 1996&            dusty &  $1000 - 2800$ & JOLA$+$Grains\\
       Allard et al. 1996&          NextGen &  $900 -  9000$ &            OS\\
      Allard et al. 1997b&    NextGen-dusty &  $900 -  3000$ &   OS$+$Grains\\   
&&&\\
       Marley et al. 1996&                  & ~~$\dots - 1000$ & K-coefficients\\
\tableline
\end{tabular}
\end{center}
\end{table}

Because they mask emergent photospheric fluxes that would otherwise
escape between absorption lines, the JOLA and SM approximations
generally led to an excessive entrapment of heat in the atmosphere
which yields systematically hotter model structures, and higher
effective temperature (T$_{\rm eff}$) estimates for individual stars.
Allard et al. (1997) have reviewed in detail the results of brown
dwarfs and VLM model atmosphere calculations with respect to the
effective temperature scale of M dwarfs.  We reproduce in Figure
\ref{Teffscale} the T$_{\rm eff} - (V-I)$ relation of Allard et al.
(1997) for the models listed in Table~\ref{grids}. 

\begin{figure}[t]
\hspace*{-1cm}
\caption[]{\label{Teffscale}Current model-dependent effective
temperature scales for cool stars down to the hydrogen burning limit.
Triangles feature results from spectral synthesis of selected stars
from the works of Kirkpatrick et al. (1993) and Leggett et al. (1996)
as indicated.  The new generation of OS models by Brett (1995b) and
Allard et al. (1996), as interpolated onto theoretical isochrones by
Chabrier et al. (1996), reproduce closely the independently-determined
positions of the eclipsing M~dwarf binary system CM Dra and YY Gem, and
the empirical T$_{\rm eff}$ scale of Jones et al. (1994).}
\end{figure}

Two double-line spectroscopic and eclipsing M dwarf binary systems, CM
Draconis and YY Geminorum, offer some guidance in the sub-solar mass
regime and are reported in Figure~\ref{Teffscale} according to Habets
\& Heintze (1981).  The use of an OS treatment of the main molecular
opacities, in particular for TiO, appears to yield a break-through in
the agreement of T$_{\rm eff}$ scales with these two M dwarfs binary
system.  The NextGen and MARCS models yield effective temperatures
that are coincidentally in good agreement with those derived
empirically from the H$_2$O opacity profile by Jones et al. (1994)
\footnote{Note that a comparison to observed spectra reveals
uncertainties of the order of 0.2 to 0.5 mag on the published $I$
magnitudes of the latest-type M dwarfs Gl406, VB10 and LHS2924
analyzed by Jones et al. (1994) and reported on Figure~\ref{Teffscale}.}.  
Note, however, that the Atlas12 OS models suffer from an inaccurate
TiO absorption profile and a complete lack of H$_2$O opacities, and
are therefore clearly inadequate in the regime of VLM stars (i.e. 
below T$_{\rm eff} \approx 4500$~K) where molecular opacities
dominate the stellar spectra and atmospheric structures.

Some uncertainties on the metallicity of the CM Draconis system may
soon disqualify the latter as a member of the disk main sequence (Viti
et al. 1997). This stresses the importance of finding other low-mass
eclipsing binary systems in the disk.  These are hopefully soon to be
provided by the 2MASS and DENIS surveys (see D. Kirkpatrick and
X. Delfosse elsewhere in this volume).  Much uncertainty remains,
therefore, at the lowermost portion of the main sequence.  The
inclusion of grain formation (as discussed below) and more complete
opacities of TiO promise a better understanding of the stars and brown
dwarfs in the vicinity of the hydrogen burning limit (the location of
which is roughly indicated in Figure~\ref{Teffscale} by the termination 
point of the Allard et al. 1996 model sequence), but still remain to be 
ascertained.

\section{The Infrared Colors of Brown Dwarfs}

\begin{figure}
\hspace*{-1cm}
\caption{\label{h2o}The observed infrared spectral distribution of the
dM8e star VB10 as obtained at UKIRT by Jones et al. (1994) (bold full
line) is compared to model spectra obtained using (from bottom to
top): (i) the SM laboratory opacity profile of Ludwig (1971), (ii) the
20 million line list by J{\o}rgensen et al. (1994), (iii) the
preliminary ab initio line list of 6.2 million transitions by Miller
\& Tennyson (1994), and (iv) the latest ab initio list of 300 million
lines by Partridge \& Schwenke (1997). The models (shown as dotted
lines) are all fully converged and normalized to the observation at
$1.2~\mu$m.  Their parameters were determined from a fit to the
optical stellar spectra (not shown) and are nearly the same in all
four cases.  Note that all 300 million lines of the Partridge \&
Schwenke list have been included in the model construction!}
\end{figure}

\begin{figure}[t]
\hspace*{-1cm}
\caption[]{\label{IJK}The most recent models of late type dwarfs are
compared to the photometric observations of field stars and brown
dwarfs, and to Pleiades objects including the brown dwarfs PPl15,
Teide1 and Calar3. Unresolved binarity is reflected in this diagram by
a red excess in $J-K$. The red dwarfs newly discovered by DENIS (see
X. Delfosse elsewhere in this volume) are also shown, although their
photometry is still very uncertain at this point.  The field brown
dwarf Gliese 229B is off the scale to the blue in $J-K$ due to strong
CH$_4$ absorption in the $K$ bandpass.  This diagram offers excellent
diagnostics to identify brown dwarf candidates of the field (very red
in either $J-K$ or $I-J$) or of the halo (very blue in both $I-J$ and
$J-K$).}
\end{figure}

The DENIS and 2MASS infrared sky surveys will soon deliver large data
bases of red dwarfs, brown dwarfs and perhaps extrasolar planets,
which will necessitate the best possible theoretical foundation.  A
proper understanding of their colors is essential in the search for
brown dwarfs.  Brown dwarfs and giant planets emit over 65 
radiation in the infrared ($>1.0 \mu$m).  Yet the main difficulties
met by VLMs and brown dwarf modelers in recent years has been to
reproduce adequately the infrared (1.4 to $2.5 \mu$m) spectral
distribution of dwarfs with spectral types later than about M6. All
models listed in the central part of table~\ref{grids} underestimate
the emergent flux, most as much as 0.5 mag at the $K$ bandpass,
despite the different opacity sources used by the authors.  Allard et
al. (1994, and subsequent publications) have explored water vapor
opacity data from various sources.  Figure~\ref{h2o} summarizes these
results.  Clearly, the water vapor opacity profile is quite uncertain
and has varied with the degree of completeness and the assumptions
used in the construction of the molecular model and its potential
surface.  The most recent and complete line list of Partridge \&
Schwenke succeeds for the first time in reproducing the $1.6~\mu$m
opacity minimum, in the $H$ bandpass, well enough for the atomic
Na$\,$I resonance line to finally emerge in the synthetic spectrum,
matching the observed feature.  However, it fails to provide an
improvement in the $K$ bandpass where the less complete list of Miller
\& Tennyson still yield the best match of the models to the observed
spectra.  The NextGen models of Allard et al. (1996) are computed
using the Miller \& Tennyson line list and are the only models to
provide a match to the infrared colors of VLMs.  This is shown in
Figure~\ref{IJK} where the complete series of NextGen models --- as
interpolated on the Baraffe et al. (1997) isochrones for 10 Gyrs and
120 Myrs and ranging from metallicities of [M/H]$= -2.0$ to 0.0 ---
are compared to the photometric field dwarfs' samples of Leggett
(1992), Tinney et al. (1993), and Kirkpatrick et al. (1995).  Other
models series including those of Brett (1995) and the Extended grid of
Allard \& Hauschildt (1995, not shown) are distinctively bluer than
the observed sequence, while the 10 Gyrs NextGen models of solar
metallicity follow closely the empirical sequence\footnote{Note that
this sequence was defined by stars selected from their optical
spectroscopic properties. The somewhat erratic aspect of the sequence
in this infrared diagram reflects uncertainties in the photometry and
perhaps in the age of the selected stars.} of Kirkpatrick \& McCarthy
(1994) until spectral types of M6 (i.e. $J-K \approx 0.85$). Beyond
this point, all models fail to reproduce the bottom of the main
sequence into the brown dwarf regime as defined by Gl406, VB10,
BRI0021 and GD165B.  The models catch up only at the much lower
T$_{\rm eff}$ of the evolved brown dwarf Gliese 229B, i.e. 900-1000~K
(Allard et al. 1996, Marley et al. 1996).

The cause of the model discrepancies at the stellar-to-brown dwarf
boundary can only be one that affects the cooler models for Gliese
229B in a far lesser obvious extent.  Since the infrared spectral
distribution is sensitive to the mixing length, yet without allowing
for an improved fit of VLMs spectra, Brett (1995a) suggested that the
problem lie in the inadequacy of the mixing length formalism for
treating the convective transport in an optically thin photospheric
medium. These concerns may also be augmented by uncertainties about
the extent of the overshooting phenomenon in VLMs (see F. D'Antona
elsewhere in this volume).  The convection zone recedes gradually
below the photosphere as the mass (and T$_{\rm eff}$) decreases along
the isochrones. This implies that the lithium test of substellarity
(Rebolo et al. 1992) --- which relies on the assumption that the brown
dwarf is still fully convective and mixing lithium from its core to
its photospheric layers after 10$^8$ yrs of age --- is inapplicable
for objects cooler than T$_{\rm eff} \leq 2200$~K. The presence of
lithium in the spectra of a late-type ($\geq $M10) field dwarfs, if
detected, could only reflect their initial abundances and {\bf not}
their substellar nature.  The shrinking of the convection zone also
allows a very good agreement between the models of Marley et
al. (which includes adiabatic convection only for the optically thick
layers of the atmosphere) and the models of Allard et al. (1996)
(based on a more careful treatment of convection with the mixing
length formalism) for the brown dwarf Gliese 229B (see Figure 5 of
Allard et al. 1997).  Yet the maximum radial extent of the convection
zone occurs at around T$_{\rm eff} = 3000\,$K, while the discrepancy
with the infrared observations increases steadily towards the bottom
of the main sequence.

A more promising answer to the so called ``infrared problem'' may
rather be found in the formation of dust grains in the very cool
(typically T$_{\rm layer} \approx$ T$_{\rm eff} - 1000\,$K) upper
layers of red and brown dwarf's atmospheres. Tsuji et al. (1996a)
proposed, based on their results of including the effects of the
formation and opacities of three grain species (Al$_2$O$_3$, Fe, and
MgSiO$_3$) in their new ``dusty'' models, that the greenhouse heating
of grain opacities, the resulting enhanced H$_2$O dissociation, and
the infrared flux redistribution can explain the infrared spectra of
cool M dwarfs.  The formation of perovskite dust grains at the expense
of TiO may also explain the observed saturation (and disappearance in
GD165B and Gliese 229B) of the TiO bands in the optical spectra of
late-type red dwarfs (see also Jones \& Tsuji elsewhere in this
volume). The implications of this result is far reaching. Field brown
dwarf candidates such as BRI0021 and GD165B can be far cooler and less
massive than previously suspected (see e.g. the NextGen-dusty model
predictions in Figure\ref{Teffscale}). If grains also form in the
young Pleiades brown dwarfs PPl15, Teide1 and Calar3 (T$_{\rm eff}
\approx $ 3000, 2800, and 2700~K respectively), lithium abundances
derived from grainless models and synthetic spectra such as those of
Pavlenko et al. (1995, see also elsewhere in this volume) may be
overestimated, and the masses attributed to these objects possibly
underestimated. Evolution models of brown dwarfs, which are sensitive
to the treatment of the atmospheres (Baraffe et al. 1995, Chabrier et
al. 1996), and their predicted Mass-lithium abundance and
Mass-Luminosity relations may also be affected.

And indeed, the temperatures and pressure conditions of the outer
layers of red dwarfs are propice to the formation of dust grains as
demonstrated years ago by Sharp \& Huebner (1990) and Burrows et
al. (1993). However it was not clear at the time if the inward
radiation of an active chromosphere, or the efficient convective
mixing from the interior, would heat up these upper photospheric
layers and disable grain formation. Another concern is that, under the
gravities prevailing in M dwarfs, gravitational settling may occur
that would eliminate large grains and their opacities from the
photospheres over relatively short time scales.  These possibilities
still need to be thoroughly investigated, but clearly, grain formation
is a process that occurs in M dwarf and brown dwarf model atmosphere
and it must included in such calculations.

In order to investigate which grains may form in the upper layers of M
dwarfs, Allard et al. (1997b, in preparation) have modified the
equation of states used in the NextGen models to include the detailed
calculation of some 1000 liquids and crystals, using the free Gibbs
energies compiled by Sharp \& Huebner. Their results showed that,
besides the three species considered by Tsuji et al., the M dwarfs
atmosphere were rich in condensates with ZrO$_2$, Ca$_2$Al$_2$SiO$_7$,
Ca$_2$MgSiO$_7$, MgAl$_2$O$_4$, Ti$_2$O$_3$, Ti$_4$O$_7$, CaTiO$_3$,
and CaSiO$_3$ showing up in models as hot as T$_{\rm eff} =
2700-3000\,$K (i.e dM8-dM6)! The preliminary NextGen-dusty models have
been computed using a continuous distribution of ellipsoid shapes and
interstellar grain sizes (between 0.025 and 0.25 $\mu$m) for the
treatment of the opacities of the Al$_2$O$_3$, Fe, MgSiO$_3$, and
Mg$_2$SiO$_4$ dust grains (see Allard \& Alexander elsewhere in this
volume for computational details). This contrast with the assumption
of spherical grains with 0.1~$\mu$m diameters in the dusty models
Tsuji et al.  Both model sets are shown in Figures~\ref{Teffscale}
and~\ref{IJK}.  As can be seen, the dusty models of Tsuji et
al. provide the correct tendency of the coolest models to get rapidly
very red (as much as $J-K = 1.65$ for GD165B) with decreasing mass for
a relatively fixed $I-J$ color. Those models are however
systematically too red in $I-J$ by as much as 1~mag and do not
reproduce even the most massive M dwarfs while over-predicting the
effects of grains in Gliese 229B type brown dwarfs (Tsuji et al.,
1996b), a problem which must be related to the use of the JOLA
treatment of molecular opacities in these models (see section
\ref{intro} above).  The NextGen-dusty models, on the other hand, show
the onset of grain formation effects by a progressive deviation from
the grainless NextGen models for $J-K \geq 0.85$, bringing an improved
agreement with the observed sequence in the region where the grainless
NextGen models deviate.  Of course, much remains to be improved in the
computation of models with dust grains. The size distribution of
various grain species, in particular those of the perovskite CaTiO$_3$
which is responsible for the depletion of TiO from the optical spectra
of late-type dwarfs (eg. GD165B, see D. Kirkpatrick elsewhere in this
volume) and of corundum (Al$_2$O$_3$) which accounts for most of the
grain opacities in current models, is unknown for the conditions
prevailing in M dwarfs atmospheres.  It is conceivable that grains
form more efficiently in M dwarfs atmospheres than in the interstellar
medium and therefore their opacities are larger than considered in the
NextGen-dusty models. We may as well be missing a number of important
contributors (e.g. ZrO$_2$) to the total grain opacities in the
models. Further investigations including time dependent grain growth
analysis will be required to determine the true contribution of dust
grains to the infrared colors of red and brown dwarfs.

In the meanwhile, diagrams like that of Figure~\ref{IJK} may help in
distinguishing interesting brown dwarfs candidates from large data
banks of detected objects, and in obtaining an appreciation of the
spectral sensitivity needed to detect new brown dwarfs. Models (Tsuji
et al. 1995, Allard et al. 1996, Marley et al. 1996) and observations
of Gliese 229B (see B. Oppenheimer elsewhere in this volume) have
shown that methane bands at 1.7, 2.4 and 3.3~$\mu$m appear in the
spectra of cool evolved brown dwarfs, and cause their $J-K$ colors to
get progressively bluer with decreasing mass and as they cool over
time. Yet their $I-J$ colors remain very red which allows to
distinguish them from hotter low-mass stars, red shifted galaxies, red
giant stars, and even from low metallicity brown dwarfs that are also
blue due to pressure-induced H$_2$ opacities in the $H-$to$-K$
bandpasses. Fortunately, grain formation and uncertainties in
molecular opacities are far reduced under low metallicity conditions
([M/H]$<-0.5$).  Therefore, model atmospheres of metal-poor subdwarf
stars and halo brown dwarfs are more reliable than their metal-rich
counterparts at this point. This has been nicely demonstrated by
Baraffe et al. (1997) who reproduced closely the main sequences of
globular clusters ranging in metallicities from [M/H]$= -2.0$ to
$-1.0$, as well as the sequence of the Monet et al. (1992) halo
subdwarfs in color-magnitude diagrams (see G. Chabrier elsewhere in
this volume).  The colors of halo brown dwarfs as predicted by the
NextGen models are therefore of quantitative quality await
confrontation with the infrared colors of metal-poor subdwarfs from
e.g. the Luyten catalog and the US Naval Observatory surveys.  The
sensitivity of the $I-J$ index to the chemical composition of the
atmosphere (clearly illustrated by the NextGen model grid) allows to
distinguish brown dwarf populations independently of an accurate
knowledge of the parallaxes or distances involved.  Even young brown
dwarfs of lower gravity appear to form a distinct sequence at bluer
$I-J$ (and redder $J-K$) values then that of their older field star
counterparts as also evident from a comparison of the 10 Gyrs and 120
Myrs NextGen models.  This gravity effect, and perhaps enhanced grain
formation, may explain the scatter of spectroscopic properties
observed among field dwarfs at the bottom of the main sequence
(Kirkpatrick, this volume), as well as the systematic differences
between Pleiades brown dwarfs and older field stars of same spectral
type (i.e. same VO band strengths) noted by Mart\`{\i}n et al. (1996).

\begin{figure}[t]
\hspace*{-1cm}
\caption[]{\label{detectBD}Predicted absolute fluxes of brown dwarfs
at 50 pc as compared to the sensitivity of ground and space-based
platforms which will be or are currently applied to the search for
brown dwarfs and extrasolar planets.  The latter are values reported
for the 5~$\sigma$ detection of a point source in 1 hr of integration,
except for the three NICMOS cameras where the integration is limited
to 40 minutes (Saumon et al. 1996).  Models of both Allard et
al. (1996) (full) and Marley et al. (1996) (dotted) are shown which
simulate (i) a brown dwarf near the hydrogen burning limit (topmost
spectrum: T$_{\rm eff}=2000$K), (ii) an evolved brown dwarf similar to
Gliese 229B (central spectra: T$_{\rm eff}=900$K and $960$K), and
(iii) a brown dwarf closer to the deuterium burning limit (lowermost
spectrum: T$_{\rm eff}= 500$K).  The corresponding black-body (dashed)
are also shown for comparison.}
\end{figure}

Gravity effects have also been found to affect the infrared spectra of
cool evolved brown dwarfs such as Gliese 229B: Allard et al. (1996)
reported a strong response of the 2.2 $\mu$m opacity minimum to
gravity changes which allowed to restrain the mass of the brown dwarf
\footnote{Only within the error on the flux calibration of the
observed spectra which are unfortunately large for this object.}.  The
general spectral distributions of cool evolved brown dwarfs are well
reproduced by current models despite the difference in their
respective modeling techniques, and despite the uncertainties tied to
grain formation and incomplete opacity data base for methane and
ammonia.  The models of Allard et al. (1996) and Marley et al. (1996)
are compared in Figure~\ref{detectBD} which also summarizes the
predicted absolute fluxes that free-floating brown dwarfs would have
at a distance of 50 pc.  As can be seen, there is no clear cut
distinction between brown dwarfs and planets; molecular bands most
gradually form (dust, H$_2$O, CH$_4$ and NH$_3$) and recede (TiO, VO,
FeH, and CO) from the stellar to the planetary regime as the
atmospheres get cooler.  They remain very bright in the $IJK$ region,
and become gradually redder in the near-infrared $I$ to $J$
bandpasses, which allows their detection from ground-based
facilities. Layers of dust clouds in their upper atmospheres may
increase the albedo of extrasolar planets and cool brown dwarfs
sufficiently to reflect the light of a close-by parent star, becoming
therefore resolvable in the optical where the clouds are densest but
the parent star is however brightest.  The peak of their intrinsic
spectral energy distribution is located at 4.5~$\mu$m.  At 5~$\mu$m,
the hotter (younger or more massive) brown dwarfs and stars show
strong CO bands which cause their flux to drop by nearly 0.5 dex
relative to that at 4.5~$\mu$m.  And between 4.5 and 10 $\mu$m,
opacities of CH$_4$ (and H$_2$O in the hotter brown dwarfs) cause the
flux to drop by 0.5 to 1.0 dex.  Searches in the 4.5-5~$\mu$m region
should therefore offer excellent possibilities of resolving brown
dwarfs and EGPs in close binary systems, and to find free-floating
brown dwarfs if space-telescope time allocations allow.  The detection
limits of current and planned ground-based and space-based telescopes
Saumon et al. (1996) are also indicated in Figure~\ref{detectBD} which
show that brown dwarfs within 50~pc would be easily detected by SIRTF
in the 4.5-5.0~$\mu$m region.  The drop in sensitivity of the various
instruments redwards of 10 $\mu$m implies, however, that brown dwarfs
and planets cooler than Gliese 299B have little chance to be detected
in those redder bandpasses.

\section{Conclusions}

In these exciting times where discoveries of brown dwarfs are finally
breaking through, model atmospheres are also rapidly becoming up to
the task of interpreting the observations and deriving new search
strategies. Uniform grids of dwarf stars and brown dwarfs model
atmospheres exist that extend from the tip to the toes of the main
sequence -- and beyond: 9000K to 900K, logg= 3.0-6.0, and [M/H]= 0.0
to $-2.0$ for the NextGen models.  These large model grids allowed the
construction of consistent interior and evolution models for VLMs that
yield unprecedent agreement with globular cluster main sequences
observed to 0.1 M$_\odot$ with HST. They led to the derivation of the
important mass-luminosity relation for halo brown dwarfs and so to the
realization that brown dwarfs cannot make up a significant fraction of
the halo missing mass.

The effective temperature scale of K to M type dwarfs with spectral
types earlier than M6 is now unambiguously established, with only
small uncertainties remaining from a possible incompleteness of
existing TiO line lists.  Grain formation has been identified as an
important process in M dwarfs and brown dwarfs atmospheres which could
explain the long-standing difficulties of the models to reproduce the
spectral distribution of dwarfs later than about M6. The results of
the models indicate that it may {\bf not} longer be assumed that the
convection zone extends to the photosphere of late-type red dwarfs and
brown dwarfs, and that their photospheric lithium abundance reflect
their core temperature and mass. The basic assumption supporting the
lithium test of substellarity is only valid for young, hot brown
dwarfs such as those found in the Pleiades cluster. Fortunately, if
the lithium test cannot identify transition objects and brown dwarfs
of the field, the OS molecular opacity treatment and grain formation
have introduce new gravity (hence age) effects in the NextGen models
that were not seen in the previous Extended models and that will
potentially allow to separate younger transitional objects from field
stars as readily as from their location in color-color diagrams. For
this the colors of late-type red dwarfs need to be known with good
accuracy i.e. better than about 0.05 magnitude, which we find is not
the case of many known late M dwarfs such as Gl406, VB10, and
especially LHS2924.

As cooler dwarfs are being discovered, spectral types are stretching
far beyond the classical Morgan \& Keenan scheme. The lack of TiO
bands in the optical, and the emergeance of CH$_4$ opacities in the
infrared in GD165B and Gl229B call for an extension of the MK system
beyond M9 to another spectral class (see D. Kirkpatrick, this volume).
While the spectral class should only reflect the effective
temperatures and not necessarily the mass of the objects, perhaps a
suitable class for these objects would nevertheless be ``T dwarfs'' as
in reminescence of J.C. Tarter who introduced the term ``brown dwarf''
now commonly accepted to designate substellar dwarfs, and Takashi
Tsuji who led the field of late dwarfs atmospheres since the early
1960's, first introduced methane as a spectral indicator of
substellarity, and who is retiring soon.  Another spectral class,
perhaps ``P'', will then be needed for dwarfs cooler then the
condensation point of water vapor including planets.  In any case,
studies of the optical spectra of Gliese 229B, GD165B, the DENIS and
2MASS objects and other late-type dwarfs will soon allow to determine
the stellar surface coverage of dust clouds if such are present, and
to verify if intrinsic spectral-type variability afflict cool dusty
dwarfs. Models will be the subject of further investigations relative
to grain formation and its effect on late-type dwarfs until they can
reproduce the lower main sequence and lead the way into the regime of
cool brown dwarfs.  Finally, if brown dwarfs are not abundant in the
halo, they certainly are in the galactic disk and their study remains
one that shall flourish as the census of the solar neighborhood
continues and the gap between planets and stars fills in.

\acknowledgments

This research is supported by a NASA LTSA NAG5-3435 and a NASA EPSCoR
grant to Wichita State University.  It was also supported in part by
NASA ATP grant NAG 5-3018 and LTSA grant NAG 5-3619 to the University
of Georgia.  Some of the calculations presented in this paper were
performed on the IBM SP2 of the UGA UCNS, at the San Diego
Supercomputer Center (SDSC) and the Cornell Theory Center (CTC), with
support from the National Science Foundation. We thank all these
institutions for a generous allocation of computer time.

\end{document}